\documentclass[prd,twocolumn,superscriptaddress]{revtex4}
\usepackage{graphicx}

\usepackage{amssymb}
\IfFileExists{url.sty}{\usepackage{url}}
                      {\newcommand{\url}{\texttt}}

\makeatletter

\providecommand{\LyX}{L\kern-.1667em\lower.25em\hbox{Y}\kern-.125emX\@}

\newcommand{\Hefour}{$^{\mathrm{4}}\mathrm{He}$\ }

\makeatother

\begin{document}
\preprint{This line only printed with preprint option}

\title{Precision Primordial $^\mathrm{\bf 4}$He measurement from the CMB}

\author{Greg Huey}
\email{greg0huey`at`isildur.astro.uiuc.edu}
\affiliation{Department of Physics, University of Illinois, Urbana, IL
61801}
\author{Richard H. Cyburt}
\email{cyburt@astro.uiuc.edu}
\affiliation{Department of Physics, University of Illinois, Urbana, IL
61801}
\author{Benjamin D. Wandelt}
\email{bwandelt@uiuc.edu}
\affiliation{Department of Physics, University of Illinois, Urbana,
IL 61801}
\affiliation{Department of Astronomy, UIUC, 1002 W Green
Street, Urbana, IL 61801}
\
\begin{abstract}
  Big bang nucleosynthesis (BBN) and the cosmic microwave background (CMB) are
two  major pillars of cosmology. Standard BBN accurately predicts the primordial
light element abundances ($^{\mathrm{4}}$He, D, $^{\mathrm{3}}$He and
$^{\mathrm{7}}$Li), depending on one parameter, the baryon density. Light
element observations are used as a baryometer. The CMB anisotropies also
contain information about the content of the universe which allows an important
consistency check on the Big Bang model. In addition CMB observations now have
sufficient accuracy to not only determine the total baryon density, but also
resolve its principal constituents, H and $^{\mathrm{4}}$He. We present a
global analysis of all recent CMB data, with special emphasis on the
concordance with BBN theory and light element observations. We find $\Omega
_{B}h^{2}=0.0250^{+0.0019}_{-0.0026}$ and $Y_{p}=0.250^{+0.010}_{-0.014}$ (fraction of
baryon mass as $^{\mathrm{4}}$He) using CMB data alone, in agreement
with \Hefour abundance observations. The determination of $Y_{p}$ allows us
to constrain the relativistic degrees of freedom during BBN, measured through
the effective number of light neutrino species, $N_{\nu,eff} = 3.02^{+0.85}_{-0.79}$, 
in accord with the Standard Model of Particle physics.  With this concordance
established we show that the inclusion of standard, $N_{\nu,eff} \equiv 3$, BBN theory priors
significantly reduces the volume of parameter space.  In this case, we find
$\Omega_{B}h^2=0.0245^{+0.0015}_{-0.0028}$ and $Y_p =
0.2493^{+0.0007}_{-0.0010}$.  We also find that the inclusion of
deuterium abundance observations reduces the $Y_p$ and $\Omega_{B}h^2$
ranges by a factor of $\sim $2. Further light element
observations and CMB anisotropy experiments will refine this concordance
and sharpen BBN and the CMB as tools for precision cosmology.
\end{abstract}
\maketitle

\section{Introduction}
Big-bang nucleosynthesis (BBN)  and
the cosmic microwave background anisotropy (CMB) are two  pillars of the
hot big bang model. 

The theory of BBN has long stood as an emblem of the predictive power
of the Big Bang model \cite{wssok,sarkar,osw,PDG_reveiw}. BBN has
long provided the most reliable measurement of the cosmological baryon
density. However, the CMB is rapidly becoming the preferred method for
determining the baryon density, with its rapidly increasing precision.
With the CMB, light element abundance observations become a powerful
probe of the universe.

The observation and analysis of cosmic microwave background (CMB)
anisotropies have attracted a great deal of attention in recent years
due to their unique relevance for cosmological theory (see
\cite{DodelsonHu} for a recent review). A flood of
observational results have been published during the last two
years\cite{Toco98,DASI,Boomerang02,Maxima,VSA,ACBAR,CBI,WMAP}. These
observations taken together measure CMB anisotropies over a large
range of angular scales. The CMB is sensitive to the
properties of the photon-baryon fluid and hence allows a precision
determination of the baryon density at redshift $z\sim1000$.

It is therefore apparent that combining  BBN and CMB provides an
opportunity for meaningful consistency checks on the standard 
cosmology and has the potential to be powerfully predictive probe of
nuclear and particle astrophysics at low and high redshift
\cite{Schramm_Turner_98,CFO2,CFO3,barger,Hansen_01,Kneller_01,DiBari_Foot_01}.  
Given the constraint on the baryon density from the CMB, BBN yields a
tight prediction of primordial $^{\mathrm{4}}\mathrm{He}$ abundance.

We also explore the promise of combining the CMB data with
measurements of the deuterium (D/H)
abundance, showing that current  measurements of (D/H) can combine with
the CMB constraints to reduce the error bars on \Hefour by another
factor of 2.

In section 2 we explain the data (CMB, BBN) and the method used to
determine the likelihood surface in parameter space - the
Metropolis-Hastings Markov Chain
Monte Carlo (MCMC) algorithm. In section 3 the parameter confidence intervals
that are extracted from that likelihood surface are discussed. In
section 4 we discuss the implications of our results and how precision
might be increased with further cosmological data.

\begin{figure*}
\includegraphics[  width=.7\textwidth,
  keepaspectratio,
  angle=0]{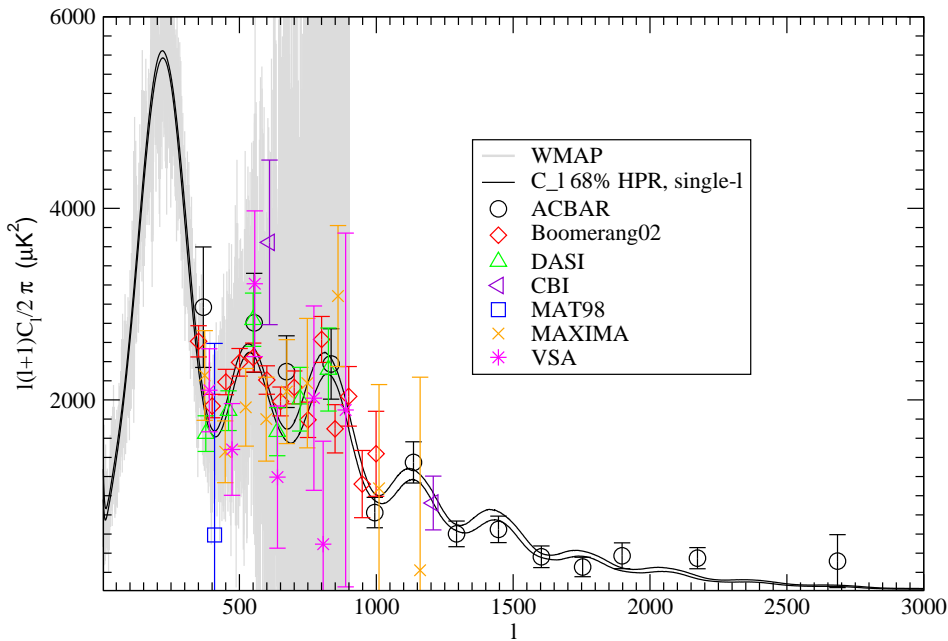}

\caption{The figure shows the data used in this paper (see
 legend) and 68\% 
 confidence intervals in the space of inferred theoretical power spectra (solid).} 
\end{figure*}

\section{methods}

Our cosmological parameters are measured by use of the Markov
Chain Metropolis Algorithm. We allow the $^{\mathrm{4}}$He mass fraction to
float as an independent variable, yielding the following parameter
space: $\Omega _{m},\Omega _{\Lambda },n_{s},h,\Omega
_{B}h^{2},Y_{p},\tau ,r,n_{t}$ 
where $\Omega _{(m,\Lambda ,B)}$ is the matter, cosmological constant,
and baryon contents, respectively, $Y_{p}$ is the fraction of the
baryons in $^{\mathrm{4}}$He by mass, $h$ is the Hubble constant in units
of $100km/sMpc$, $n_{(s,t)}$ are the power-law index of the primordial
scalar and tensor perturbations respectively, $r$ is the fraction
of the observed CMB quadrupole that is tensor and $\tau $ is the
optical depth to the last-scattering surface (that is, re-scattering
of CMB photons by reionization is allowed for). Note that in our selection
of the parameter space simplifying assumptions have been made: adiabatic,
scale-free primordial perturbations, the universe contains only cold
dark matter and a cosmological constant and the neutrino species are
strictly those of the Standard Model. To determine the region of this
parameter space allowed by experimental data, one must sample the
space over a wide range of points. At a given point the relative likelihood
of the parameter values yielding the observations must be determined,
and the range of points sampled must adequately cover the space. Our
primary likelihood calculation is a comparison of the simulated CMB
spectra produced by CMBFAST~\cite{CMBFAST} against the WMAP CMB
experiment~\cite{WMAP}, along with smaller-scale (bin $\ell _{eff}>350$)
data from the following experiments: Toco98~\cite{Toco98}, DASI~\cite{DASI},
Maxima~\cite{Maxima}, VSA~\cite{VSA}, ACBAR~\cite{ACBAR},
%
%
Boomerang02~\cite{Boomerang02}, and CBI~\cite{CBI}~\footnote{For WMAP, the 
published likelihood function was used. For the other experiments BJK 
formalism was used, with the BJK parameters obtained from either RADPACK, 
or the collaborations directly.}. We include the published calibration 
uncertainties for each experiment and find the maximum likelihood value 
for these parameters at each point in the cosmological parameter space.

Figure 1 shows the data we used as
well as the most likely 68\% of our inferred theory power spectra.

One could attempt to sample the parameter space on a uniform grid,
but the high dimensionality coupled with the computational demands
of CMBFAST makes this impossible in a reasonable amount of time. 
Instead, we implemented the Metropolis-Hastings MCMC Algorithm\cite{Metropolis,Knox,
BridleLewis}: starting from the current
point in parameter space $\overrightarrow{X}_{i}$ one proposes
a test point $\overrightarrow{p}$, drawn from a distribution described
by a density function called the proposal
density: $P(\overrightarrow{p}|\overrightarrow{X}_{i})$. The choice of
proposal density is somewhat arbitrary, but a poor choice will cause
the parameter estimation procedure to be inefficient.
%
The likelihood relative to $\overrightarrow{X}_{i}$ is computed at the test
point $\overrightarrow{p}$.
If the likelihood at $\overrightarrow{p}$
is greater, then point $\overrightarrow{X}_{i+1}=\overrightarrow{p}$.
Otherwise there is a probability that $\overrightarrow{X}_{i+1}=\overrightarrow{p}$
equal to the likelihood of $\overrightarrow{p}$ divided by the likelihood
of $\overrightarrow{X}_{i}$. If the proposal density is not symmetric
in its arguments then this probability needs to be corrected by the
factor
$P(\overrightarrow{X}_{i}|\overrightarrow{p})/P(\overrightarrow{p}|\overrightarrow{X}_{i})$
in order to enforce detailed balance.
If $\overrightarrow{X}_{i+1}$ is not
set to the the point $\overrightarrow{p}$, then it is set to the
current point: $\overrightarrow{X}_{i+1}=\overrightarrow{X}_{i}$.
After a sufficient number of iterations the 
%
resulting density of points $\left\{ \overrightarrow{X}_{i}\: |\: i=1\ldots n\right\}$
asymptotically approaches the likelihood function
on parameter space. What constitutes a {}``sufficient'' number of points
is in general difficult to determine (and impossible to determine
with absolute certainty) - though there are tests which are good indicators.
We use a test for our Markov Chain which was suggested by
\cite{Gelman92} and also used by the WMAP team\cite{WMAP_Verde}.

We tune our Markov Chain code for optimal efficiency by first finding
the maximal likelihood point in parameter space using a global
maximization method (within our prior space, see below) then using it as
the starting point for a sample Markov Chain. The variance of the
sample chain is used to compute a step size matrix that will be used
by the main chains. An efficient Markov Chain should take steps that
are not too large or too small - either will result in an inefficient,
slowly converging chain. 

We choose our proposal density to be a multivariate Gaussian, whose
covariance matrix  $V_{s}$ is proportional to the covariance of the
sample chain $V_{a}$.
The proportionality constant $\alpha (D)$ is chosen
so if the underlying distribution were Gaussian, $50\%$ of the Markov
chain points would be accepted. Its value depends on the number of
dimensions of parameter space, $D$ (for us $D=9$). We found:\begin{equation}
\alpha \left(D\right)\simeq 0.54784\, D-0.36159\label{eq:var_ration_half_accept}\end{equation}

We ran two sets of Markov Chains, $A$ and $B$, each consisting of
$20$ independent chains. Each chain was started at a point chosen
from the distribution of the sample chain. Set $A$ had only very
weak top-hat priors ($\Omega_{B}h^{2}\in \left[0.014,0.030\right],$
$Y_{p}\in \left[0.13,0.34\right]$, $h\in \left[0.45,0.95\right]$,
$\Omega_{m}\in \left[0.03,1.00\right]$,
$\Omega_{\Lambda}\in \left[0.00,0.97\right]$,
$n_{s}\in \left[0.5,1.4\right]$,
$n_{t}\in \left[-3.0,3.0\right]$,
$r\in \left[0.0,3.0\right]$,
$\tau\in \left[0.0,1.0\right]$),
%
%
whereas set $B$ additionally had two strong priors: a BBN
consistency condition between $\Omega _{B}h^{2}$ and $Y_{p}$~\cite{CFO1},
and the constraint $h=0.74+0.11-0.094$~\cite{Hproj}~\footnote{The strong prior 
$h=0.74+0.11-0.094$ is an asymmetric Gaussian with plus and minus sigmas as given, 
and represents the constraint coming from the Hubble key project~\cite{Hproj}}.
%
%
The BBN consistency condition is simply this: for a given
baryon density one expects BBN to produce a certain abundance of $^{\mathrm{4}}$He,
with some theoretical error (mostly driven by uncertainties in the
nuclear cross-sections). In set $A$ we treat $\Omega _{B}h^{2}$
and $Y_{p}$ as two independent variables. In set $B$ we enforce
theoretical self-consistency between those variables.
This non-CMB constraint is incorporated in a Bayesian way. The
additional information is included as an additional prior and the
density to be sampled from (the posterior density) becomes 
\begin{equation}
\begin{array}{c}
P_{tot}\left(\overrightarrow{X}\right) =  P_{CMB}\left(\overrightarrow{X}\right) P_{BBN}\left(\Omega_{B}h^{2}, Y_p\right) P_{Hubble}\left(h\right)\\
= P\left(\overrightarrow{X}|CMB\right)
P\left(Y_p|\Omega_{B}h^{2}\right) P\left(h|Hubble\;data\right) 
\end{array}
\label{eq:BBNprobexp_1}\end{equation}
where $P\left(x|y\right)$ is the conditional density for getting $x$,
given $y$. Note that $P\left(Y_p|\Omega_{B}h^{2}\right)$ is purely a
theoretical prior enforcing the BBN relation, and contains no
abundance measurements. The Markov Chain then automatically explores
the new, more constrained region of parameter space.
 As one might
expect, sets $A$ and $B$ differ significantly in their parameter
space coverage, and thus their proposal densities and chain starting
points were determined separately.

Though a Markov Chain approach saves significant computational time,
it is difficult to guarantee after some number of points that the
chain has converged sufficiently to the true, underlying distribution.
Indeed, a chain cannot tell one anything about a region
it has not yet visited. We use a convergence test suggested by Gelman
\& Rubin~\cite{Gelman92}, which was also employed by
WMAP~\cite{WMAP_Verde}.

We have generalized this criterion to multiple dimensions, keeping
in mind that any convergence test must be covariant (if a transformation
of parameter space can change the determination of {}``convergence'',
then the test is a bad one). Each chain out of the $20$ has
its own mean and variance~\footnote{For $D>1$ we use the word variance
 to mean covariance matrix.}. If each chain reflected the underlying
distribution, then the variance of the means of the chains should
be much less than the variance of the underlying distribution. We
thus compute the variance of the chain means, multiply with the inverse
of the average chain variance, and take the trace:\begin{equation}
\begin{array}{c}
 U\equiv \frac{1}{N-1}{\displaystyle \sum _{j=1}^{N}\left(\overrightarrow{\overline{X}}_{j}-\overrightarrow{\overline{X}}\right)\otimes \left(\overrightarrow{\overline{X}}_{j}-\overrightarrow{\overline{X}}\right)}\\
 W\equiv \frac{1}{N}{\displaystyle \sum _{j=1}^{N}\frac{1}{n_{j}-1}}{\displaystyle \sum _{i=1}^{n_{j}}\left(\overrightarrow{X}_{j,i}-\overrightarrow{\overline{X}}_{j}\right)\otimes \left(\overrightarrow{X}_{j,i}-\overrightarrow{\overline{X}}_{j}\right)}\\
 \mu \equiv Tr\left[UW^{-1}\right] / D\end{array}
\label{eq:cnvgtest_1}\end{equation}
where $N$ is the number of chains (20), $n_{j}$ is the number of
points in chain $j$, $\overrightarrow{\overline{X}}$ is the total
mean, $\overrightarrow{\overline{X}}_{j}$ is the mean of chain $j$.
We require that $\mu <0.1$. Set $B$ easily satisfies this criteria
with $30,000$ points, whereas set $A$ required about $60,000$ points.
The average chain variance, $W$, is used because this underestimates
the variance of the distribution until convergence is attained.

As a self-consistency check, one can take the point distribution of
set $A$ and combine it with the BBN $\Omega _{B}h^{2}-Y_{p}$ and
Hubble Key project priors. The resulting distribution should be the
same as set $B$. The extent to which these distributions differ is
a measure of non-convergence of the sets. We determined that the $68\%$
confidence regions of these distributions more than $95\%$ overlap
in the $\left(\Omega _{B}h^{2},Y_{p}\right)$ plane. A Markov Chain
can be combined with a prior after the generation of the chain by
assigning a weight to each point. The likelihood of a region in parameter
space is then the weighted density of the points in that region. Because
a Markov Chain maintains the full $D$-dimensional likelihood distribution
in the parameter space, after it is generated the chain may be convolved
with any arbitrary other likelihood function in that parameter space.
Thus one can generate a Markov Chain distribution for WMAP alone,
and chose any subset of the other cosmological datasets to convolve
it with - for very little additional CPU cost. This is the basis of
the \emph{Cosmic Concordance Project} (CCP) web-site
\url{http://galadriel.astro.uiuc.edu/ccp/}  where the parameter
constraints from chains $A$ and $B$ can be explored and combined with
other cosmological datasets and priors. Further details about the  CCP
and about our parameter estimation methodology will be given in \cite{CCP}.

\begin{figure*}
\includegraphics[  width=.7\textwidth,
  keepaspectratio,
  angle=270]{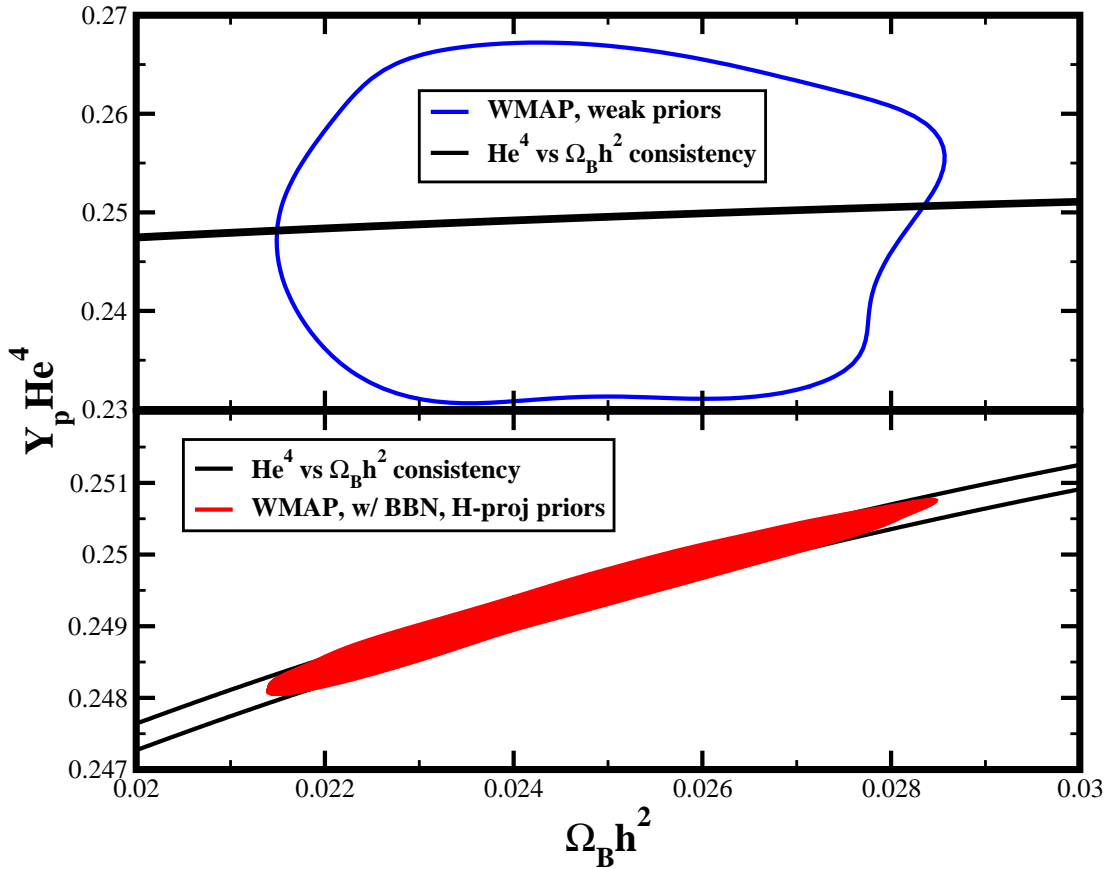}

\caption{The top figure shows the $68\%$ confidence region in the
  $\Omega _{B}h^{2}$, $Y_{p}$ plane
from the CMB data alone (WMAP and high-$\ell $ data from other recent
CMB experiments) bounded by the blue line. The black line bounds the
$68\%$ 
confidence region from BBN theory alone, using the best fit $\Omega
_{B}h^{2}$ from the CMB. Note that the BBN theory
band is in good agreement with the CMB data.
The bottom figure shows the result of combining
CMB and BBN data using an expanded $Y_{P}$ axis. The solid red region
is the $68\%$ confidence region 
for the set $B$ Markov chains which have as priors the BBN constraint
and the Hubble Key project constraint on $h$. Note that the BBN constraint
greatly reduces the allow range of $Y_{P}$ as a function of $\Omega_{B}h^{2}$.}

\end{figure*}

\begin{figure*}
\includegraphics[  width=.7\textwidth,
  keepaspectratio,
  angle=270,
  origin=lB]{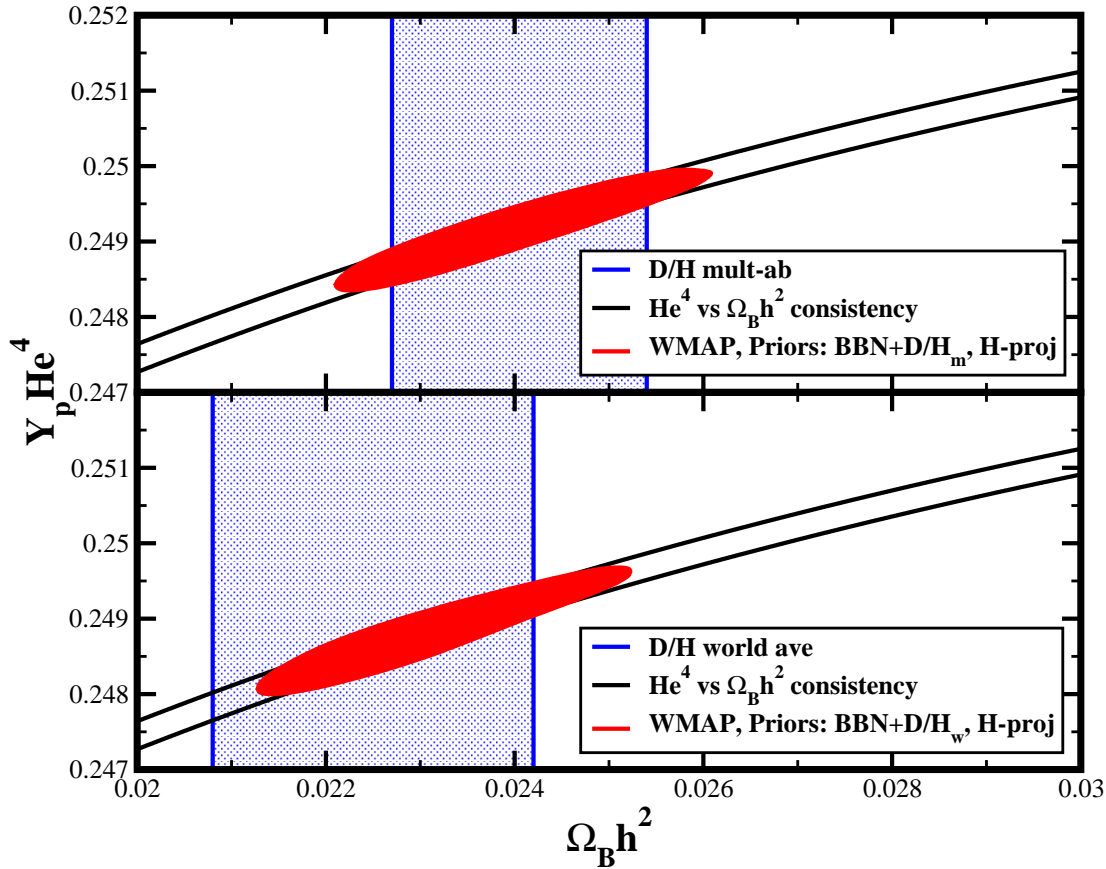}

\caption{The primordial abundance of deuterium can be used to further constrain
the baryon density (blue shaded band is $68\%$), which in turn increases
the precision of the $Y_{P}$ determination (solid red region is $68\%$).
Two different $D$ abundances are used: 2a) data from 2 multiple-line
absorption systems, and 2b) data from 5 systems (the 2 multiple-line
plus 3 others) (see text).}
\end{figure*}

\section{Results}

As our Markov Chain sets $A$ and $B$ approached $60,000$ and $30,000$
points respectively, convergence test eq.~\ref{eq:cnvgtest_1} gave
$\mu \backsim 0.05-0.06$~%
\footnote{For this value we concatenated $5$ chains into $1$, thus transforming $20$
chains per set to $4$.%
} and we declared our chains sufficiently converged to provide reliable
statistics. Figure~2a shows the 68\% confidence region of set $A$
in the $\left(\Omega _{B}h^{2},Y_{p}\right)$ plane.  Thus from CMB
data alone we find $\Omega _{B}h^{2}=0.025^{+0.0019}_{-0.0026}$ 
and $Y_{p}=0.250^{+0.010}_{-0.014}$

It is worth noting also that Markov chain set A yields a tight constraint on
the neutrino number during BBN: Allowing the number of
BBN neutrinos to float, one can find a 68\%
confidence interval for the number of BBN neutrinos needed to yield the
determined $Y_p$ from the determined $\Omega_{B}h^{2}$. Given a BBN
code that computes the probability density
$P(Y_p|\Omega_{B}h^{2},N_{\nu,eff})$ (where the stochasticity is due
to the measurement uncertainty in the relevant nuclear cross-sections)
we can compute $P(N_{\nu,eff})$ based on CMB data and BBN theory as 
\begin{eqnarray}
P(N_{\nu,eff}|\mathrm{ BBN,CMB})&=&\nonumber\\\int dY_p d(\Omega_{B}h^{2})
P(N_{\nu,eff}|Y_p,\Omega_{B}h^{2}) P(Y_p,\Omega_{B}h^{2})&=&\nonumber\\
\int dY_p d(\Omega_{B}h^{2})P(Y_p|\Omega_{B}h^{2},N_{\nu,eff})P(Y_p,\Omega_{B}h^{2}).\nonumber
\end{eqnarray}
The first term under each integral  enforces the BBN relation and the second
term the CMB posterior. The second equality holds true if we assume flat priors
$P(Y_p|\Omega_{B}h^{2},BBN)=const.$ and
$P(N_{\nu,eff}|\Omega_{B}h^{2},BBN)=const$ for the range of parameter
space of interest \footnote{ For simplicity, we approximate
$P(Y_p,\Omega_{B}h^{2}|CMB)=P(Y_p|CMB) P(\Omega_{B}h^{2}|CMB)$ for
this calculation. We also assume that the CMB does not constrain
$N_{\nu,eff}$ (instead, we treat the effective number of neutrino
species that are relevant for the CMB anisotropy as independent of
$N_{\nu,eff}$) and that there is no {\it a priori} preference for any
value of $N_{\nu,eff}$ or $Y_p$ for any value of $\Omega_{B}h^{2}$. }. We find
$N_{\nu ,eff}=3.02^{+0.85}_{-0.79}$.  $N_{\nu,eff}$ is consistent with the
standard model value of 3 and previous studies 
\cite{CFO2,CFO3,barger,Hansen_01,Kneller_01,DiBari_Foot_01}.

However, $\Omega _{B}h^{2}$ and $Y_{P}$
are jointly constrained by BBN theory, and thus are not really independent
variables. Adopting the standard BBN model ($N_{\nu,eff}({\rm BBN})$=3) of~\cite{CFO1} yields a consistency
relation between $\Omega _{B}h^{2}$ and $Y_{P}$. The $68\%$ confidence
region of this consistency relation appears in Figure~2a as a narrow
band (narrow enough that the upper and lower bounding curves appear
to merge). Enforcing this condition greatly increases the precision
of parameter estimation, as evident in table~1, with
dramatic affect on $Y_{P}$ measurement:
$Y_{P}=0.2493^{+0.0007}_{-0.0010}$. This is simply a result of the
accuracy with which \Hefour is determined by BBN ($\sim 0.1\%$).  In
Figure~2b we have zoomed in on this CMB-BBN concordance region. Also
shown is the $68\%$ confidence region of set $B$ shaded as solid
red. As one would expect, set $B$ agrees with the product of the
CMB (set $A$) and BBN (consistency band) likelihoods. It is important
to note that in Figure~2a the agreement between the CMB and BBN allowed
regions need not have happened. Instead, the BBN consistency band
might not have passed through the high CMB likelihood region, which
would have forced one to consider a BBN scenario other than the standard
model one with  3 neutrinos. The CMB-BBN agreement reaffirms
the standard BBN scenario.

Model A and Model B compare quite well to \Hefour
observations. Olive, Skillman and Steigman
(1997)~\cite{Olive_Skillman_Steigman_97} and Fields and Olive
(1998)~\cite{Fields_Olive_98} find $Y_p = 0.238 \pm 0.002$, while
Izotov and Thuan (1998)~\cite{Izotov_Thuan_98} find $Y_p = 0.244 \pm
0.002$.  The errors cited are statistical only.  Comparing these
numbers, not only are they discrepant from each other, but they lie
below the mean value determined in this evaluation.  However, Olive
and Skillman (2001)~\cite{Olive_Skillman_16} critically evaluate the
methods used in determining $Y_p$ and find a lower bound to a
systematic error, $\sigma_{sys} \geq 0.005$.  This systematic error is
added in quadrature with the statistical error to determine the total
error, increasing the errors to $0.0054$.  The $Y_p$ observations
are brought into marginal accord with each other and the CMB; both
lie systematically lower than the CMB determined value.  As
discussed earlier, the systematic error used is only a lower bound,
and as such the true errors are most likely larger than those quoted.

\begin{table}
\label{tab:numb}
\caption{Cosmological Parameter Estimates}
\begin{center}
\begin{tabular}[t]{|c|c|c|}
\hline
Parameters & Model A & Model B \\ \hline\hline
$\Omega_{B}h^2$ & $0.0250^{+0.0019}_{-0.0026}$ &
$0.0245^{+0.0015}_{-0.0029}$ \\ \hline
$Y_p$ & $0.250^{+0.010}_{-0.014}$ & $0.2493^{+0.0007}_{-0.0010}$ \\
\hline\hline 
$h$ & $0.684^{+0.057}_{-0.097}$ & $0.733\pm 0.059$ \\ \hline
$\Omega_{M}$ & $0.241^{+0.062}_{-0.064}$ & $0.219^{+0.041}_{-0.058}$ \\
\hline
$\Omega_{\Lambda}$ & $0.792^{+0.063}_{-0.047}$ &
$0.798^{+0.060}_{-0.045}$ \\ \hline
$n_s$ & $1.047^{+0.062}_{-0.075}$ & $1.024^{+0.041}_{-0.086}$ \\ \hline
$r$ & $0.168^{+0.065}_{-0.141}$ & $0.117^{+0.084}_{-0.117}$ \\ \hline
$n_t$ & $0.089^{+0.456}_{-0.258}$ & $0.207^{+0.408}_{-0.347}$ \\ \hline
$\tau$ & $0.228^{+0.103}_{-0.123}$ & $0.180^{+0.058}_{-0.127}$ \\
\hline\hline
\end{tabular}
\end{center}
\end{table}

In Figure~2 the allowed $\Omega _{B}h^{2}$ is very large. Any other
data that can reduce the allowed $\Omega _{B}h^{2}$ range will have
the additional benefit of refining the precision of the $Y_{P}$ measurement.
As an example, we consider $D$ abundance in Figure~3. The value
of $D/H$ is still a somewhat open question due to low number
statistics, and thus we demonstrate 
the effects of two different $D$ abundances. In Figure~3a we use
the average of the 2 multiple absorption line systems
of~\cite{OMeara01_Dma1,Kirkman03_Dma2}: $D/H=\left(2.49\pm 0.18\right)\times 10^{-5}$. The blue shaded band
is the $68\%$ confidence baryon density range allowed by this $D/H$
value as determined with the BBN theory of~\cite{CFO1}. The solid red
region is the $68\%$ confidence region resulted 
from the convolution of the $D$ baryon range and the concordance
region of Figure~2b. Thus we find $Y_{p}=0.2491^{+0.0004}_{-0.0005}$,
$\Omega _{B}h^{2}=0.0237^{+0.0010}_{-0.0012}$. Alternatively, one can
use a conservative combination of 5 $D/H$ measurements, including the
two multiple 
absorption line systems employed above~\cite{OMeara01_Dma1,Kirkman03_Dma2,Burles98a_Dwa1,Burles98b_Dwa2,Pettini01_Dwa3}:
$D/H=\left(2.78\pm 0.29\right)\times 10^{-5}$ (the overall error
increase is because the other three systems are not consistent with
each other or the multiple absorption line systems, a hint of an
underlying systematic error for the single absorption line
systems). For this $D$ abundance we find
$Y_{p}=0.2488^{+0.0004}_{-0.0005}$ and $\Omega
_{B}h^{2}=0.0230^{+0.0008}_{-0.0012}$.

\section{Conclusions}
This work has been based on two general ideas; (1) BBN and the CMB
independently probe two different epochs, providing valuable consistency
checks for the underpinnings of the Standard Cosmological
Model; (2) having established that the cosmological model agrees
remarkably well with these very different observational probes we use
these data to to make a precision measurement of the $^\mathrm{4}$He abundance.  We have presented an analysis of all recent CMB data, in which
we have determined the cosmic baryon density and the primordial helium
abundance. We found $\Omega _{B}h^{2}=0.0250^{+0.0019}_{-0.0026}$ and
$Y_{p}=0.250^{+0.010}_{-0.014}$ at $68\%$ from CMB data alone. This is
consistent with 3 standard model neutrinos during BBN.

We have shown that this is fully consistent with the predicted $^{\mathrm{4}}$He
abundance from BBN theory, and marginally consistent with
$^{\mathrm{4}}$He observations. The likely source of this slight
discrepancy is an underestimate of the dominant, systematic
uncertainties in the $^{\mathrm{4}}$He observations, which now seems
affirmed with the CMB determination of $Y_{p}$. The agreement between the
CMB-only set $A$ confidence region of $\Omega _{B}h^{2}$, $Y_{p}$
and the consistency band based on BBN theory shown in Figure~2a reaffirms
the standard BBN model. Thus using BBN theory, we can effectively
remove $Y_{p}$ as a freely floating variable, enforcing the $\Omega
_{B}h^{2}$-$Y_{p}$ BBN relation in CMB data analysis. 
Given this, we found the incorporation of BBN theory into parameter
extraction from CMB data results in a precision measurement of the
\Hefour abundance. We find
$\Omega_{B}h^2=0.0245^{+0.0015}_{-0.0029}$ and $Y_p =0.2493^{+0.0007}_{-0.0010}$. 

Using this CMB-BBN determined $Y_p$, one can study
and constrain stellar
evolution~\cite{Bono_02,Cassisi_Salaris_Irwin_03}.  One can similarly study the 
nucleosynthetic history of all the light element 
abundances as discussed in~\cite{CFO3} and references therein.

We show the promise of incorporating deuterium abundance observations, yielding
$Y_{p}=0.2491^{+0.0004}_{-0.0005}$, $\Omega
_{B}h^{2}=0.0237^{+0.0010}_{-0.0012}$ or $Y_{p}=0.2488^{+0.0004}_{-0.0005}$,
$\Omega _{B}h^{2}=0.0230^{+0.0008}_{-0.0012}$ depending on which systems
are used to measure the deuterium abundance.

The addition of the $D$ abundance observations are only one example
of many possible cosmological datasets that might be incorporated
into parameter extraction to increase precision. An experiment may
constrain a parameter directly, or may reduce degeneracy in a related
parameter. For example, using large-scale structure information to
reduce the residual $n_{s}$-$\Omega _{B}h^{2}$ degeneracy in the
current CMB data would also increase the precision of the $Y_{P}$
determination. Also, further light element observations and CMB
anisotropy experiments will refine this concordance and sharpen BBN
and the CMB as tools for precision cosmology. Due to the effect of the \Hefour abundance on the damping tail,
this may improve the constraint on a possible running of scalar spectral index.   These are the topics of
on-going work and can be further  explored at the \emph{Cosmic
Concordance Project} web-site:
\url{http://galadriel.astro.uiuc.edu/ccp/} \cite{CCP}. 

Many thanks to Prof.~Brian Fields for helpful discussions.
The work of R.H.C was provided by the National Science Foundation
grant AST-0092939. This work was partially supported by National Computational
Science Alliance under grant AST020003N and utilized the NCSA Platinum
system. 

\end{document}